\documentclass[aps,prl,reprint, twocolumn,superscriptaddress]{revtex4-2}
\usepackage{blindtext}
\usepackage[colorlinks=true,allcolors=blue]{hyperref}

\usepackage{blindtext}
\usepackage{bm}

\def \beq{\begin{equation}}
\def \eeq{\end{equation}}

\usepackage{amsmath}
\usepackage{amssymb}
\usepackage{graphicx}
\usepackage{graphicx,xcolor}

\graphicspath{{figs}}
\usepackage[capitalize]{cleveref}
\crefname{equation}{Eq.}{}
\crefrangeformat{equation}{Eqs.~(#3#1#4)-(#5#2#6)}

\begin{document}

\title{Active phase separation from concentration-dependent diffusion in chemical networks}

\author{Giulia Pisegna}
\email{giulia.pisegna@gmail.com}
\affiliation{Max Planck Institute for Dynamics and Self-Organization (MPI-DS), D-37077 Göttingen, Germany}
\author{Ramin Golestanian}
\email{ramin.golestanian@ds.mpg.de}
\affiliation{Max Planck Institute for Dynamics and Self-Organization (MPI-DS), D-37077 Göttingen, Germany}
\affiliation{Rudolf Peierls Centre for Theoretical Physics, University of Oxford, Oxford OX1 3PU, United Kingdom}

\begin{abstract}
Active phase separation in living systems is ubiquitous, yet the microscopic origins of the non-equilibrium terms in active field theories remain largely unknown. Here, starting from stochastic mass-conserving reaction-diffusion dynamics of two interacting chemical species, we perform a systematic slow-mode reduction and derive the complete Active Model B+ field theory, with all phenomenological coefficients expressed in terms of microscopic parameters. We identify concentration-dependent diffusivities and asymmetric inter-species chemotactic couplings as the minimal microscopic ingredients generating the genuinely non-equilibrium coefficient responsible for microphase separation, cluster phases, and arrested coarsening. Our framework recovers the equilibrium limit consistently and is validated by stochastic numerical simulations. These results establish a direct, bottom-up link between the architecture of chemical reaction networks and the universality class of their emergent large-scale dynamics.
\end{abstract}

\maketitle

Spatial self-organization in living systems emerges from the interplay of chemical reactions, diffusion, and non-equilibrium transport. Reaction-diffusion models have provided a minimal framework for understanding how non-linear reactions coupled to diffusion produce non-equilibrium patterns \cite{turing_chemical_1952, 
prigogine1968symmetry, halatek2018rethinking, brauns2020phase, Cotton2022, 
AgudoCanalejo2018, AgudoCanalejo2020}, while many phenomena of intracellular condensation, such as nucleoli \cite{boisvert2007multifunctional, 
lafontaine2021nucleolus} and P-granules \cite{brangwynne2009germline}, have been  rationalized within phenomenological theories of active phase separation 
\cite{hyman2014liquid, weber2019physics}. Despite their common goal, these two approaches have largely evolved independently, and a systematic connection between them remains lacking.

When chemical reactions preserve the total molecular mass, the dynamics can be described by mass-conserving reaction-diffusion models (McRD), which underlie intracellular protein pattern formation such as the MinDE system in bacteria 
\cite{halatek2012highly, halatek2018rethinking, chardac2025mind} and have been widely studied in multispecies systems \cite{brauns2020phase, brauns2021wavelength}. 
At the field-theoretical level, conserved dynamics provides the natural setting for active phase separation, described by Active Model B$+$ (AMB$+$) 
\cite{cates2025active, tjhung2018cluster, fausti2021capillary}. Recent 
progress has been done connecting McRD to active field theories \cite{bergmann2018active, 
bergmann2019system, robinson2025universal, toffenetti2026active,ilker2026hydrodynamic}, yet how stochastic 
fluctuations propagate from micro to meso-scales has not been systematically addressed. Crucially, the microscopic origin of the non-equilibrium and non-integrable $\zeta$ coefficient of AMB$+$, responsible for microphase separation and arrested coarsening, remains to be clarified.

The goal of this Letter is to establish a direct bridge between microscopic reaction-diffusion dynamics and active field theories by systematically coarse-graining spatially extended chemical processes into continuum descriptions with phenomenological coefficients expressed in terms of microscopic parameters. 
Here, we consider a stochastic dynamics with a non-equilibrium state-dependent mobility matrix, encoding non-linear diffusivities and inter-species phoretic couplings \cite{Golestanian2022a}. The assumption of constant mobility is, in fact, often unrealistic in biological media, where transport occurs in heterogeneous cellular environments and is strongly influenced by biochemical interactions \cite{roussel2004reaction, AgudoCanalejo2020}. 
While concentration-dependent diffusivities have been investigated in reaction-diffusion models \cite{polanowski2006reaction, roussel2004pattern}, their consequences for mass-conserving systems remain unexplored. We show that asymmetric 
cross-diffusion is the minimal microscopic ingredient generating $\zeta \neq 0$, and that stochastic currents are dominated by diffusive rather than reactive 
fluctuations.

\begin{figure}[b]
    \centering
    \includegraphics[width=\linewidth]{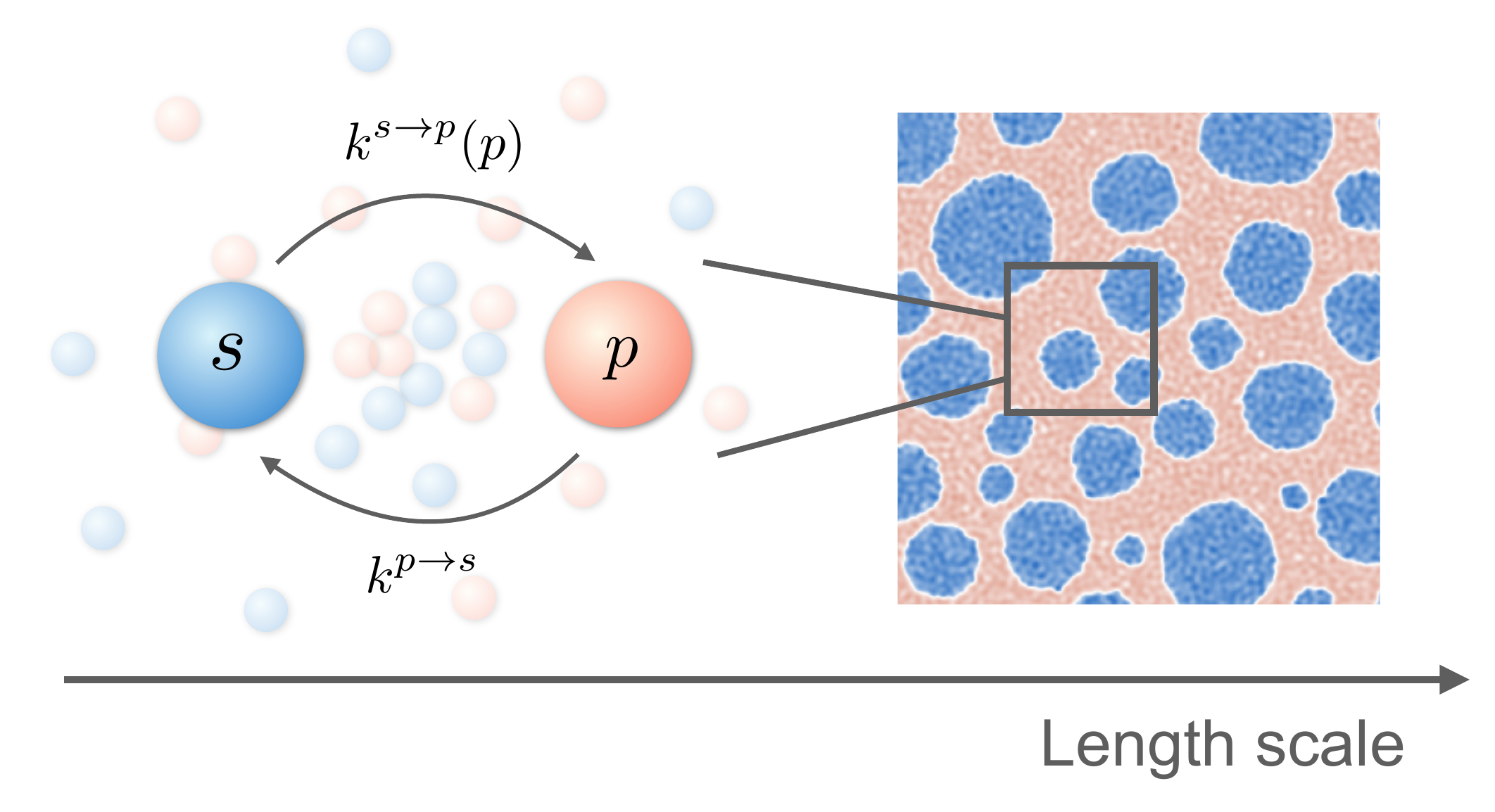}
    \caption{\textbf{Coarse-graining McRD leads to active field theories.} The number of chemical species and associated conservation laws determines the emergent coarse-grained dynamics. A two-species system with non-linear interconversion between substrate $s$ and product $p$ possesses a single conserved field and maps onto AMB$+$ dynamics, leading to active phase separation.}
    \label{fig:schematic}
\end{figure}

\textit{McRD with concentration-dependent mobility.}
We consider a stochastic reaction-diffusion model for $N$ chemical species of concentrations $ u^a(\bm x,t)$, whose chemical interactions are encoded in reactive terms $ R^a(\bm u)$, where $\bm u = (u^1, u^2,...,u^N)^T$.  Including spatial transport, the dynamics read,
\beq 
\label{eq:model2}
\partial_t u^a = \bm \nabla \cdot \Big[ \sum_b M^{ab}(\bm u) \bm \nabla \mu^b \Big] + R^a + \xi^a + \bm \nabla \cdot \bm \Xi^a .
\eeq 
The chemical potential $\mu^b = \delta \mathcal F/ \delta u^b$ derives from the free energy of an ideal gas $\mathcal F = \int_{\bm r} \sum_a [ \mu_0^a u^a + k_{\rm B} T u^a ( \ln u^a-1)]$. The system responds to this thermodynamic force with a mobility matrix $M^{ab}(\bm u)$. Chemical reactions produce a reactive force  $R^a(\bm u) =  S^a [\omega_+(\bm u) - \omega_-(\bm u)]$, where $S^a$ is the stoichiometric coefficient and $\omega_\pm$ are the forward and 
backward reaction fluxes. The conservation of total mass requires $\sum_a R^a = \sum_a S^a=0$. 

 Eq.\eqref{eq:model2} is thermodynamically consistent provided: (i) the mobility $M^{ab}$ is a positive, semi-definite, symmetric matrix, namely satisfying Onsager reciprocal relations \cite{de2013non}, (ii) the reactive fluxes fulfill $k_{\rm B} T \ln ( \omega_+ /\omega_-) = - \sum_a \mu^a S^a$  \cite{avanzini2024nonequilibrium}.
 Both conditions are satisfied by linear reaction fluxes and by the mobility $M^{ab} = (k_{\rm B} T)^{-1} (L^a u^b \delta^{ab}-L^{ab} u^a u^b)$, with constant coefficients and symmetric $L^{ab}$ \cite{akaberian2023nonequilibrium}.
 As shown in \cite{SM}, this functional form arises naturally from a microscopic lattice-jump process, which also yields the stochastic fluctuations with zero mean and variance,
\begin{align}
\label{eq:noisezeta}
  \langle \xi^a (\bm x,t) \xi^b(\bm x',t') \rangle & =  f^{ab}(\bm u) \delta (\bm x-\bm x') \delta(t-t'),  \\
  \label{eq:noisecsi}
\langle \Xi_i^a (\bm x,t) \Xi_j^b(\bm x',t') \rangle &=  2 k_{\rm B} T \delta_{ij} M^{ab}(\bm u) \delta (\bm x-\bm x') \delta(t-t') .
\end{align}
Both non-conservative  and  conservative multiplicative noises are present, the latter typical of diffusive systems \cite{gardiner2009stochastic, dean1996langevin}. Reactions instead generate $\xi^a$ whose amplitude depends on the stoichiometry of the reactions \cite{SM}. For a two-species mass-conserving case, the reactive noise amplitude reads 
$f^{ab}(\bm u) = [\omega_+ + \omega_-](\delta^{ab} - \sigma^{1,ab})$, 
with $\sigma^1$ the first Pauli matrix \cite{SM}.

As formulated above, the model satisfies equilibrium conditions, but it can be driven out of equilibrium in two ways:  (i) by promoting the mobility coefficients to concentration-dependent functions, $L^a, L^{ab} \to L^a(\bm u), 
L^{ab}(\bm u)$, while keeping those of the noise amplitude constant, thereby violating the fluctuation-dissipation relation, or  (ii) through non-linear 
reaction fluxes \cite{turing_chemical_1952}. Either route admits a description in terms of an active field theory at large scales. To connect with the standard formulation of reaction-diffusion models, we recast the diffusive dynamics in $\partial_t  u^a = \bm \nabla \cdot ( \sum_b  D^{ab}(\bm u) \bm \nabla u^b) +  R^a + \text{noise}$,
with diffusivity matrix,
\beq 
\label{eq:diffusivity}
D^{ab}(\bm u) = M^{ab}(\bm u) k_{\rm B} T/ u^b.
\eeq 
The minimal conditions on $D^{ab}$ and $ R^a$ for AMB$+$ behavior to emerge  at large scales are established below.

\textit{Two chemical species and AMB+.} We focus on a two-species system composed of a substrate $s(\bm x,t)$ and a product $p(\bm x,t)$, interconverting through activation and degradation processes (Fig.\ref{fig:schematic}). To capture non-linear reactions,  
we consider positive feedback such that the conversion rate from substrate to product $k^{s\to p}$ increases with the local concentration of $p$, yielding the reaction term $ R^p=-R^s=R$ with $R(p,s)= k^{s \to p}(p) s - k^{p \to s}p$. A concrete realization is the wave-pinning model for cell polarization \cite{mori2008wave, robinson2025universal}, with
\beq 
\label{eq:cellpolarization}
R(p, s) = \left( k+ (1-k) \frac{p^2}{1+p^2} \right) s - p ,
\eeq 
and $k \in [0,1]$. This non-linear feedback breaks condition (ii) introduced above, and the resulting reactivity sets the noise amplitude to $f^{ab}(\bm u) = [ k^{s \to p}(p) s+k^{p \to s} p]( \delta^{ab}- \sigma^{1,ab})$. 
The total density $\rho=p+s$ is conserved and obeys, $\partial_t \rho = - \bm \nabla \cdot \bm J_\rho$, with current
\beq 
\bm J_\rho = - \sum_{a,b}\Big[ D^{ab}(\bm u) \bm \nabla u^b + \bm \Xi^a \Big ] .
\eeq
Crucially, only diffusive noise enters the conserved current: the non-conservative reactive noise cancels exactly due to mass conservation (see \cite{SM}).

As expected, conserved quantities generate slow hydrodynamic modes whose relaxation time diverges with system size. To identify them, we first consider homogeneous steady states satisfying $R^a(\bm u_0)=0$, which define the reactive nullclines in concentration space. Pattern-forming instabilities arise when the nullcline is regressive 
\cite{brauns2020phase, robinson2025universal}, reflecting an effective bistability 
required for phase separation. We thus identify the nature of the modes from the linearized dynamics of fluctuations around these states. Secondly, we adiabatically eliminate the fast mode order by order in a gradient expansion, and finally we expand to third order in the fields to obtain the nonlinear corrections that determine the AMB$+$ coefficients.

Linearizing Eq.~\eqref{eq:model2} with 
\eqref{eq:diffusivity} around $\bm u = \bm u_0 + \delta\bm u$ gives
\beq 
\label{eq:lineardeltau}
\partial_t \delta u^a = \sum_b [ D_0^{ab} \nabla^2 \delta  u^b + J^a_b \delta  u^b] +  \xi_0^a + \bm \nabla \cdot \bm \Xi_0^a,
\eeq 
with the constant matrix $D_0^{ab} = D^{ab} (\bm u_0)$, and the Jacobian defined as  $J^{a}_b = (\partial  R^a/\partial u^b )|_{\bm u_0}$. We retain only the additive part of the noise sources \eqref{eq:noisezeta} and \eqref{eq:noisecsi} evaluated at $\bm u_0$, leading to $\xi_0^a$ and $\bm \Xi_0^a$ respectively. Hereafter, subscript indices on denote partial 
derivatives with respect to the corresponding field, evaluated at $\bm u_0$.
The eigenstructure of $ J^a_b$  separates the dynamics into slow and fast modes. The latter relaxes with a finite eigenvalue $\sigma = R_p - R_s<0$ when  $R_p< R_s$. 
Projecting onto the left eigenvectors $\bm w_1=(1,1)$ and 
$\bm w_2=(R_p,R_s)/(R_p-R_s)$ of $J^a_b$ yields the dynamics for the slow mode $\phi=\rho-\rho_0$ and the fast mode $\eta$ as follows
\begin{align}
 \partial_t \phi &=  w_1^a \partial_t \delta u^a  =  w_1^a D_0^{ab} \nabla^2 \delta u^b +  w_1^a \bm \nabla \cdot \bm \Xi_0^a, \\ \nonumber 
  \partial_t \eta &=   w_2^a  \partial_t  \delta   u^a =  (R_p-R_s) \eta +  w_2^a  D_0^{ab} \nabla^2 \delta u^b +   w_2^a  \xi_0^a.
\end{align}
Inverting the linear relation, $\delta \bm u = \mathcal{W}^{-1}(\phi, \eta)^T$, with $\mathcal{W}= (\bm w_1, \bm w_2)^T$, we obtain
\begin{align}
 \partial_ t \phi &= r_0 \nabla^2 \phi + r_1 \nabla^2 \eta + \bm \nabla \cdot \tilde{\bm \Xi}_0, \\ \nonumber
 \partial_ t \eta &= (R_p-R_s) \eta + a_0 \nabla^2 \phi + a_1 \nabla^2 \eta + \tilde \xi_0, 
\end{align}
where, for $D^{sp}=D^{ps}=0$, we have
\begin{align}
 r_0 &= \frac{R_p D_0^{ss} - R_s D_0^{pp}}{(R_p-R_s)}, \quad
 r_1 = D_0^{pp} - D_0^{ss} ,  \\ \nonumber
  a_0 &= \frac{R_s R_p}{(R_p-R_s)^2}(D_0^{ss} - D_0^{pp}), \quad
 a_1 = \frac{R_p D_0^{pp} - R_s D_0^{ss}}{(R_p-R_s)} .
\end{align}
The term $r_0$ plays the role of a tuning parameter for pattern formation that requires $r_0 < 0 \rightarrow R_p/R_s > D_0^{pp}/D_0^{ss} $.
When $(R_p-R_s) <0$, $\eta$ relaxes rapidly compared with $\phi$ and can be eliminated under a quasi-stationary approximation. Retaining the leading contribution in a gradient expansion scheme, gives
$
\eta \simeq  - a_0/(R_p-R_s)  \nabla^2 \phi  
$, which can be used to close the equation for $\phi$ as follows
\beq 
\label{eq:linearphi}
 \partial_t \phi = r_0 \nabla^2 \phi - K \nabla^4 \phi + \bm \nabla \cdot \tilde{\bm \Xi}_0,
\eeq
with $K = - (D_0^{pp}- D_0^{ss})^2 (R_s R_p)/(R_p-R_s)^3 $. The noise from the enslaving is $\mathcal{O}(\nabla^4)$ and subdominant; the relevant noise amplitude is instead
\begin{align}
   &\langle \tilde{\Xi}_{0,i}(\bm x,t) \tilde{\Xi}_{0,j}(\bm x',t)) \rangle = \\ \nonumber
   & 2 k_{\rm B} T \delta_{ij} \sum_{a,b}M^{ab}(\bm u_0) \delta(\bm x- \bm x') \delta (t-t').  
\end{align}

To derive the relevant nonlinear terms, we expand Eq.~\eqref{eq:model2} to third order around $\bm u_0$, project onto the slow and fast modes as defined by $\bm w_1$ and $\bm w_2$, and systematically eliminate the fast variable. Expanding Eq.~\eqref{eq:model2} yields
    \begin{align}
    \label{eq:non-linearu}
    \nonumber
     \partial_t \delta  u^a &= \bm \nabla \cdot \Big[ \sum_b \Big( D_0^{ab} + \sum_c  D^{ab}_c \delta  u^c \\
     & + \frac12 \sum_{c,d} D^{ab}_{cd} \delta  u^c \delta u^d \Big)\bm \nabla \delta u^b \Big ] + \sum_b J^a_b \delta u^b \\ \nonumber
    & +  Q^a +  W^a +  \xi^a_0 + \bm \nabla \cdot \bm \Xi_0^a .
\end{align}
The reactive nonlinearities are found as follows
\begin{equation}
\label{eq:QW}
Q = \frac12 \sum_{a,b} R_{ab} \delta u^a \delta u^b, \quad W = \sum_{a,b,c} \frac16 R_{abc} \delta u^a  \delta u^b  \delta u^c,
\end{equation}
with $\bm Q=(Q,-Q)$ and $\bm W=(W,-W)$. Their antisymmetric structure ensures 
that $ w_1^a Q^a =  w_1^a W^a = 0$: reactive nonlinearities do not contribute directly to $\phi$, but enter through the elimination of $\eta$. The diffusivity expansion instead contributes directly to the slow mode:
\begin{align} 
\label{eq:nonlinearphi}
 \partial_t \phi &= r_0 \nabla^2 \phi  + r_1 \nabla^2 \eta + \bm \nabla \cdot \tilde {\bm \Xi}_0\\ \nonumber
&+  \bm \nabla \cdot [ b_0 \phi \bm \nabla \phi + b_1 \eta \bm \nabla \phi + b_2 \phi \bm \nabla \eta + b_3 \phi^2 \bm \nabla \phi]. 
\end{align}
Note that we discard terms of order $\sim \eta \bm \nabla \eta$ in the current because they will be irrelevant under power counting. Reactive nonlinearities instead enter indirectly through the dynamics of the fast mode with $Q(\eta, \phi)$ and $ W(\eta, \phi)$.
Expanding these terms, the dynamics of the fast mode becomes,
\begin{align}
\label{eq:nonlineta}
 \partial_t \eta &= (R_p-R_s) \eta + a_0 \nabla^2 \phi + a_1 \nabla^2 \eta + b_4 \nabla^2 \phi^2 \\ \nonumber
& + a_2 \phi^2 + a_3 \phi^3 + a_4 \eta \phi .
\end{align} 
We therefore obtain the coupled nonlinear equations for the slow Eq. \eqref{eq:nonlinearphi} and fast modes Eq. \eqref{eq:nonlineta}. The expressions of the coefficients of Eqs. \eqref{eq:nonlinearphi} and \eqref{eq:nonlineta} in terms of microscopic parameters are reported in the End Matter. In the quasi-stationary regime we can solve for $\eta$ as follows 
\begin{align}
\label{eq:etafin}
\nonumber
 \eta &\simeq  - [(R_p-R_s) + a_1 \nabla^2]^{-1}(a_0 \nabla^2 \phi + b_4 \nabla^2 \phi^2 + Q + W) \\  \nonumber
& \simeq - \frac{1}{(R_p-R_s)} \Bigl[ a_0 \nabla^2 \phi + a_2 \phi^2 + \left( a_3 - \frac{a_2 a_4}{(R_p-R_s)} \right)\phi^3 \\ 
& + \left( b_4- \frac{a_1 a_2}{(R_p-R_s)}\right) \nabla^2 \phi^2 \Bigr] .
\end{align}
The local terms in $\eta$ renormalize the effective chemical potential of $\phi$, while gradient terms generate the active contributions of AMB$+$.

\emph{Effective chemical potential.} 
Upon substituting Eq.~\eqref{eq:etafin} into 
Eq.~\eqref{eq:nonlinearphi}, we obtain a first part of the dynamics that takes the conserved form 
$\partial_t\phi = \nabla^2\mu$ where
\begin{align}
\mu = r_0\phi + \tau_0\phi^2 + u_0\phi^3 - K\nabla^2\phi ,
\end{align}
and
\begin{align}
\tau_0 &= \frac{b_0}{2} - \frac{a_2 r_1}{(R_p-R_s)} , \\
 u_0 &= \frac{1}{3}\Bigl[b_3 - \frac{a_2 (b_1+2b_2)}{(R_p-R_s)}\Bigr] - 
 \frac{r_1}{(R_p-R_s)}\Bigl(a_3 - \frac{a_2 a_4}{(R_p-R_s)}\Bigr) .
\end{align}
These coefficients arise from the interplay between nonlinear diffusion and reactive nonlinearities. 

\begin{figure}
    \centering
\includegraphics[width=1.0\linewidth]{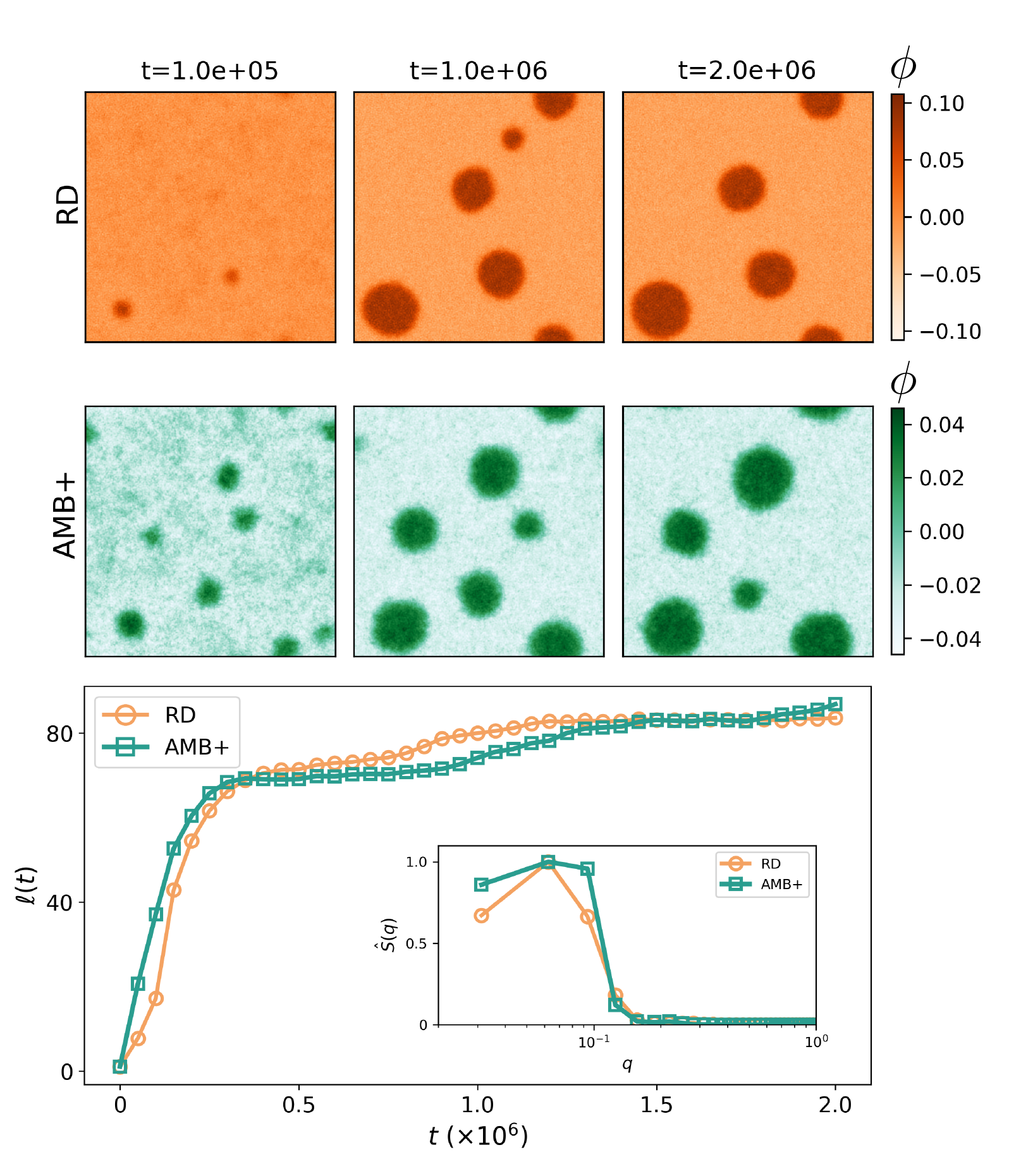}
    \caption{\textbf{Stochastic numerical simulations}. Top: snapshots of the slow mode evolution of the RD and AMB+ trajectories starting from the same initial conditions. Bottom: coarsening length scale $\ell(t)$ in time, inset: static structure factors normalized by their maximum value computed at $t=2\times10^6$. 
    Parameters in common: $N=512$, $L=64\pi$, $dt=0.025$. RD: $k=0.07,  p_0=0.5, D_0^{ss}=0.1, D_0^{pp}=0.062, D_s^{pp} =0.12, D_p^{ss}=0.216, D_n=10^{-6}, \phi_0=-0.005$. AMB$+$:
    $r_0 = -0.0040, \tau_0 =-0.3805, u_0 =12.6,K =0.073, \lambda =-0.027, \zeta=-0.205, \kappa=-1.62, D_n = 2\times 10^{-6}, \phi_0=-0.005$.   The latter parameter measure the distance from the nullcline. Note that, for this set of parameters, a homogeneous state with this values of $\phi_0$ is expected to be stable. On the other hand, noise triggers pattern formation. We used only additive noise in both models since the multiplicative part is not fundamental for the mapping.}
    \label{fig:simulations}
\end{figure}

\emph{Active terms.} Gradient corrections from the elimination of $\eta$ yield  the complete AMB$+$ dynamics. Substituting these terms into Eq.~\eqref{eq:nonlinearphi}, we finally obtain
\begin{align}
\label{eq:AMBplus}
    \partial_ t \phi &= \nabla^2 \mu + \nabla^2 [ \lambda |\bm \nabla \phi|^2 + \kappa \phi \nabla^2 \phi] \\ \nonumber 
    & - \zeta \bm \nabla \cdot [ \bm \nabla \phi \nabla^2 \phi]+ \bm \nabla \cdot \tilde{\bm \Xi}_0,
\end{align}
where the coarse-grained coefficients of activity read
\begin{align}
\nonumber
 \lambda &= \frac{2 r_1}{(R_p-R_s)}\left( \frac{a_1 a_2}{(R_p-R_s)} - b_4 \right), \\
\label{eq:activeparameters}
 \kappa &= \lambda - \frac{b_2 a_0}{(R_p-R_s)}, \\ \nonumber
 \zeta &= \frac{a_0}{(R_p-R_s)}(b_1 - b_2) . 
\end{align}
The three coefficients correspond to the characteristic active elements identified in Refs.~\cite{tjhung2018cluster}. The first term reproduces the nonlinear gradient contribution present in Active Model B, which has also been derived in related reaction-diffusion frameworks \cite{robinson2025universal}. The second term can in fact be absorbed into a field-dependent surface tension through the renormalization $K(\phi) = K - \kappa \phi$. Although both of these contributions break detailed balance, and can generate interesting non-equilibrium patterns \cite{toffenetti2026active, ilker2026hydrodynamic}, their phenomenology remains qualitatively close to that of equilibrium phase separation when fluctuations' amplitude are small preserving $K(\phi)>0$. The coefficient $\zeta$ is instead the hallmark of genuine non-equilibrium behavior, driving cluster phases, bubbly phase separation, and arrested coarsening \cite{cates2025active, tjhung2018cluster}. In terms of microscopic parameters we find,
\beq 
\zeta = \frac{R_s R_p}{(R_p-R_s)^3}(D_0^{ss}-D_0^{pp}) (D_p^{ss}-D_s^{pp}) . 
\eeq 
This is the central result of this work.
It shows that $\zeta\neq 0$ requires both different bulk diffusivities 
($D_0^{ss}\neq D_0^{pp}$) and asymmetric cross-concentration dependence ($D_p^{ss}\neq D_s^{pp}$): it is not sufficient for each species to sense only its own concentration. An analogous contribution from phoretic cross-couplings gives $\zeta\sim D_p^{ps}-D_s^{sp}$, providing an alternative microscopic route; 
both contributions add when $ D^{ab}$ has all non-zero entries.


\textit{Consistent equilibrium limit.}  The absence of concentration dependent diffusivity and non-linear phoretic interactions yields $\zeta=0$. However, fully restoring an equilibrium dynamics requires satisfying the conditions of thermodynamic consistency for the diffusivity matrix, namely
 $D^{pp}= L^p$, $D^{ss}=L^s$, $D^{ps} = - L^{ps}p$ and $D^{sp}=-L^{ps}s$. In addition, choosing a constant production rate for $p$ restores the flux condition and fixes the homogeneous equilibrium state to $s_0 = - p_0 R_p/R_s$. 
As a consequence,  $D^{ps}_p = D^{sp}_s$, $D^{ps}_s=D^{sp}_p=0$, and $D^{sp}_0= - (R_p/R_s) D^{ps}_0$. Inserting these relations in Eq.\eqref{eq:activeparameters} yields $\kappa= 2\lambda$. In this case the gradient contributions of Eq.\eqref{eq:AMBplus} can be written in a variational form coming from $
\mathcal{F}_1 = -\lambda \int d\bm{r} \ \phi |\nabla \phi|^2,
$
confirming compatibility with equilibrium thermodynamics. Finally,
$
r_0 =(D_0^{ss} + D_0^{pp})/2 > 0,
$
reflects stability of the homogeneous equilibrium state in absence of interactions. 

\textit{Numerical simulations.}  To validate the coarse-graining procedure, we  perform stochastic simulations of both the McRD model, with reactive force Eq.~\eqref{eq:cellpolarization}, and the corresponding AMB$+$ model in two dimensions. For a given parameter set of the RD model, the AMB$+$ coefficients are computed directly from our analytical expressions, with no free parameters. Fig.~\ref{fig:simulations} shows snapshots of the slow-mode field $\phi$ from both models starting from identical initial conditions: the spatial structures and their temporal evolution are in close qualitative agreement, confirming that 
the AMB$+$ dynamics faithfully captures the large-scale behavior of the 
microscopic RD system. The coarsening length scale $\ell(t)$ and the static 
structure factor $S(k)$ (Fig.~\ref{fig:simulations}, bottom) further support this agreement quantitatively. The chosen parameter set lies close to the phase boundary, resulting in slow coarsening dynamics consistent with the small values of $r_0$ obtained from the mapping. For this reason, we cannot determine whether the pattern ultimately arrests at a characteristic length scale or remains metastable, but we can use this case to assess the validity of the analysis. Since all coefficients in the dynamics of $\phi$ depend on the parameters of the underlying microscopic model, a comprehensive exploration of the parameter space and its phases remains challenging.

\textit{Conclusions.} Starting from the stochastic microscopic dynamics of two 
reacting species with nonlinear kinetics and concentration-dependent mobilities, we have performed a controlled slow-mode reduction around a nullcline steady state and derived the complete AMB$+$ dynamics, with all phenomenological coefficients 
expressed in terms of microscopic parameters and the equilibrium limit consistently recovered. Beyond providing a microscopic foundation for AMB$+$, our derivation uniquely 
identifies concentration-dependent diffusivities with asymmetric cross-species coupling as the minimal microscopic ingredient generating the non-equilibrium $\zeta$ coefficient, directly connecting the architecture of chemical networks to their large-scale self-organization.
 Our expansion captures active terms up to two fields and four gradients; a 
complete $\mathcal{O}(\nabla^4)$ expansion would straightforwardly recover 
additional active terms recently proposed in \cite{burekovic2026active},
providing a microscopic interpretation for those as well that could be relevant in different parameters regimes.

A natural extension is to generalize the framework to reaction networks with 
more than two chemical species, where multiple conserved slow modes are expected to generate richer active field theories with additional dynamical couplings and 
symmetry structures. The bottom-up methodology developed here provides a 
systematic route for deriving active continuum theories directly from 
experimentally realizable molecular systems, bridging the gap between 
phenomenological active field theories and chemically driven living and synthetic matter.
\newline

While this manuscript was in preparation, a related study appeared online, deriving similar results in a model of chemically active emulsions \cite{ilker2026hydrodynamic}. Its findings are consistent with ours and provide an independent microscopic realization of this phenomenology.

\acknowledgements
G.P. acknowledges Victor Chardès for useful discussions. This work has received support from the Max Planck School Matter to Life and the MaxSyn-Bio Consortium, which are jointly funded by the Federal Ministry of Education and Research (BMBF) of Germany, and the Max Planck Society. G.P. was supported by the Deutsche Forschungsgemeinschaft (DFG, German Research Foundation) through a Walter Benjamin fellowship PI 2211/1.
\newline

\bibliography{biblio,Golestanian}

\begin{thebibliography}{36}%
\makeatletter
\providecommand \@ifxundefined [1]{%
 \@ifx{#1\undefined}
}%
\providecommand \@ifnum [1]{%
 \ifnum #1\expandafter \@firstoftwo
 \else \expandafter \@secondoftwo
 \fi
}%
\providecommand \@ifx [1]{%
 \ifx #1\expandafter \@firstoftwo
 \else \expandafter \@secondoftwo
 \fi
}%
\providecommand \natexlab [1]{#1}%
\providecommand \enquote  [1]{``#1''}%
\providecommand \bibnamefont  [1]{#1}%
\providecommand \bibfnamefont [1]{#1}%
\providecommand \citenamefont [1]{#1}%
\providecommand \href@noop [0]{\@secondoftwo}%
\providecommand \href [0]{\begingroup \@sanitize@url \@href}%
\providecommand \@href[1]{\@@startlink{#1}\@@href}%
\providecommand \@@href[1]{\endgroup#1\@@endlink}%
\providecommand \@sanitize@url [0]{\catcode `\\12\catcode `\$12\catcode
  `\&12\catcode `\#12\catcode `\^12\catcode `\_12\catcode `\%12\relax}%
\providecommand \@@startlink[1]{}%
\providecommand \@@endlink[0]{}%
\providecommand \url  [0]{\begingroup\@sanitize@url \@url }%
\providecommand \@url [1]{\endgroup\@href {#1}{\urlprefix }}%
\providecommand \urlprefix  [0]{URL }%
\providecommand \Eprint [0]{\href }%
\providecommand \doibase [0]{https://doi.org/}%
\providecommand \selectlanguage [0]{\@gobble}%
\providecommand \bibinfo  [0]{\@secondoftwo}%
\providecommand \bibfield  [0]{\@secondoftwo}%
\providecommand \translation [1]{[#1]}%
\providecommand \BibitemOpen [0]{}%
\providecommand \bibitemStop [0]{}%
\providecommand \bibitemNoStop [0]{.\EOS\space}%
\providecommand \EOS [0]{\spacefactor3000\relax}%
\providecommand \BibitemShut  [1]{\csname bibitem#1\endcsname}%
\let\auto@bib@innerbib\@empty
\bibitem [{\citenamefont {Turing}(1952)}]{turing_chemical_1952}%
  \BibitemOpen
  \bibfield  {author} {\bibinfo {author} {\bibfnamefont {A.~M.}\ \bibnamefont
  {Turing}},\ }\bibfield  {title} {\bibinfo {title} {The chemical basis of
  morphogenesis},\ }\href {https://doi.org/10.1098/rstb.1952.0012} {\bibfield
  {journal} {\bibinfo  {journal} {Philos. Trans. R. Soc. B}\ }\textbf {\bibinfo
  {volume} {237}},\ \bibinfo {pages} {37} (\bibinfo {year} {1952})}\BibitemShut
  {NoStop}%
\bibitem [{\citenamefont {Prigogine}\ and\ \citenamefont
  {Lefever}(1968)}]{prigogine1968symmetry}%
  \BibitemOpen
  \bibfield  {author} {\bibinfo {author} {\bibfnamefont {I.}~\bibnamefont
  {Prigogine}}\ and\ \bibinfo {author} {\bibfnamefont {R.}~\bibnamefont
  {Lefever}},\ }\bibfield  {title} {\bibinfo {title} {Symmetry breaking
  instabilities in dissipative systems. ii},\ }\href
  {https://doi.org/https://doi.org/10.1063/1.1668896} {\bibfield  {journal}
  {\bibinfo  {journal} {J. Chem. Phys.}\ }\textbf {\bibinfo {volume} {48}},\
  \bibinfo {pages} {1695} (\bibinfo {year} {1968})}\BibitemShut {NoStop}%
\bibitem [{\citenamefont {Halatek}\ and\ \citenamefont
  {Frey}(2018)}]{halatek2018rethinking}%
  \BibitemOpen
  \bibfield  {author} {\bibinfo {author} {\bibfnamefont {J.}~\bibnamefont
  {Halatek}}\ and\ \bibinfo {author} {\bibfnamefont {E.}~\bibnamefont {Frey}},\
  }\bibfield  {title} {\bibinfo {title} {Rethinking pattern formation in
  reaction--diffusion systems},\ }\href
  {https://doi.org/https://doi.org/10.1038/s41567-017-0040-5} {\bibfield
  {journal} {\bibinfo  {journal} {Nat. Phys.}\ }\textbf {\bibinfo {volume}
  {14}},\ \bibinfo {pages} {507} (\bibinfo {year} {2018})}\BibitemShut
  {NoStop}%
\bibitem [{\citenamefont {Brauns}\ \emph {et~al.}(2020)\citenamefont {Brauns},
  \citenamefont {Halatek},\ and\ \citenamefont {Frey}}]{brauns2020phase}%
  \BibitemOpen
  \bibfield  {author} {\bibinfo {author} {\bibfnamefont {F.}~\bibnamefont
  {Brauns}}, \bibinfo {author} {\bibfnamefont {J.}~\bibnamefont {Halatek}},\
  and\ \bibinfo {author} {\bibfnamefont {E.}~\bibnamefont {Frey}},\ }\bibfield
  {title} {\bibinfo {title} {Phase-space geometry of mass-conserving
  reaction-diffusion dynamics},\ }\href
  {https://doi.org/https://doi.org/10.1103/PhysRevX.10.041036} {\bibfield
  {journal} {\bibinfo  {journal} {Phys. Rev. X}\ }\textbf {\bibinfo {volume}
  {10}},\ \bibinfo {pages} {041036} (\bibinfo {year} {2020})}\BibitemShut
  {NoStop}%
\bibitem [{\citenamefont {Cotton}\ \emph {et~al.}(2022)\citenamefont {Cotton},
  \citenamefont {Golestanian},\ and\ \citenamefont
  {Agudo-Canalejo}}]{Cotton2022}%
  \BibitemOpen
  \bibfield  {author} {\bibinfo {author} {\bibfnamefont {M.}~\bibnamefont
  {Cotton}}, \bibinfo {author} {\bibfnamefont {R.}~\bibnamefont
  {Golestanian}},\ and\ \bibinfo {author} {\bibfnamefont {J.}~\bibnamefont
  {Agudo-Canalejo}},\ }\bibfield  {title} {\bibinfo {title} {Catalysis-induced
  phase separation and autoregulation of enzymatic activity},\ }\href
  {https://doi.org/10.1103/PhysRevLett.129.158101} {\bibfield  {journal}
  {\bibinfo  {journal} {Phys. Rev. Lett.}\ }\textbf {\bibinfo {volume} {129}},\
  \bibinfo {eid} {158101} (\bibinfo {year} {2022})}\BibitemShut {NoStop}%
\bibitem [{\citenamefont {Agudo-Canalejo}\ \emph {et~al.}(2018)\citenamefont
  {Agudo-Canalejo}, \citenamefont {Illien},\ and\ \citenamefont
  {Golestanian}}]{AgudoCanalejo2018}%
  \BibitemOpen
  \bibfield  {author} {\bibinfo {author} {\bibfnamefont {J.}~\bibnamefont
  {Agudo-Canalejo}}, \bibinfo {author} {\bibfnamefont {P.}~\bibnamefont
  {Illien}},\ and\ \bibinfo {author} {\bibfnamefont {R.}~\bibnamefont
  {Golestanian}},\ }\bibfield  {title} {\bibinfo {title} {Phoresis and enhanced
  diffusion compete in enzyme chemotaxis},\ }\href
  {https://doi.org/10.1021/acs.nanolett.8b00717} {\bibfield  {journal}
  {\bibinfo  {journal} {Nano Lett.}\ }\textbf {\bibinfo {volume} {18}},\
  \bibinfo {pages} {2711} (\bibinfo {year} {2018})}\BibitemShut {NoStop}%
\bibitem [{\citenamefont {Agudo-Canalejo}\ \emph {et~al.}(2020)\citenamefont
  {Agudo-Canalejo}, \citenamefont {Illien},\ and\ \citenamefont
  {Golestanian}}]{AgudoCanalejo2020}%
  \BibitemOpen
  \bibfield  {author} {\bibinfo {author} {\bibfnamefont {J.}~\bibnamefont
  {Agudo-Canalejo}}, \bibinfo {author} {\bibfnamefont {P.}~\bibnamefont
  {Illien}},\ and\ \bibinfo {author} {\bibfnamefont {R.}~\bibnamefont
  {Golestanian}},\ }\bibfield  {title} {\bibinfo {title} {Cooperatively
  enhanced reactivity and {\textquotedblleft}stabilitaxis{\textquotedblright}
  of dissociating oligomeric proteins},\ }\href
  {https://doi.org/10.1073/pnas.1919635117} {\bibfield  {journal} {\bibinfo
  {journal} {Proc. Natl. Acad. Sci.}\ }\textbf {\bibinfo {volume} {117}},\
  \bibinfo {pages} {11894} (\bibinfo {year} {2020})}\BibitemShut {NoStop}%
\bibitem [{\citenamefont {Boisvert}\ \emph {et~al.}(2007)\citenamefont
  {Boisvert}, \citenamefont {Van~Koningsbruggen}, \citenamefont
  {Navascu{\'e}s},\ and\ \citenamefont {Lamond}}]{boisvert2007multifunctional}%
  \BibitemOpen
  \bibfield  {author} {\bibinfo {author} {\bibfnamefont {F.-M.}\ \bibnamefont
  {Boisvert}}, \bibinfo {author} {\bibfnamefont {S.}~\bibnamefont
  {Van~Koningsbruggen}}, \bibinfo {author} {\bibfnamefont {J.}~\bibnamefont
  {Navascu{\'e}s}},\ and\ \bibinfo {author} {\bibfnamefont {A.~I.}\
  \bibnamefont {Lamond}},\ }\bibfield  {title} {\bibinfo {title} {The
  multifunctional nucleolus},\ }\href
  {https://doi.org/https://doi.org/10.1038/nrm2184} {\bibfield  {journal}
  {\bibinfo  {journal} {Nat. Rev. Mol. Cell Biol.}\ }\textbf {\bibinfo {volume}
  {8}},\ \bibinfo {pages} {574} (\bibinfo {year} {2007})}\BibitemShut {NoStop}%
\bibitem [{\citenamefont {Lafontaine}\ \emph {et~al.}(2021)\citenamefont
  {Lafontaine}, \citenamefont {Riback}, \citenamefont {Bascetin},\ and\
  \citenamefont {Brangwynne}}]{lafontaine2021nucleolus}%
  \BibitemOpen
  \bibfield  {author} {\bibinfo {author} {\bibfnamefont {D.~L.}\ \bibnamefont
  {Lafontaine}}, \bibinfo {author} {\bibfnamefont {J.~A.}\ \bibnamefont
  {Riback}}, \bibinfo {author} {\bibfnamefont {R.}~\bibnamefont {Bascetin}},\
  and\ \bibinfo {author} {\bibfnamefont {C.~P.}\ \bibnamefont {Brangwynne}},\
  }\bibfield  {title} {\bibinfo {title} {The nucleolus as a multiphase liquid
  condensate},\ }\href
  {https://doi.org/https://doi.org/10.1038/s41580-020-0272-6} {\bibfield
  {journal} {\bibinfo  {journal} {Nat. Rev. Mol. Cell Biol.}\ }\textbf
  {\bibinfo {volume} {22}},\ \bibinfo {pages} {165} (\bibinfo {year}
  {2021})}\BibitemShut {NoStop}%
\bibitem [{\citenamefont {Brangwynne}\ \emph {et~al.}(2009)\citenamefont
  {Brangwynne}, \citenamefont {Eckmann}, \citenamefont {Courson}, \citenamefont
  {Rybarska}, \citenamefont {Hoege}, \citenamefont {Gharakhani}, \citenamefont
  {J{\"u}licher},\ and\ \citenamefont {Hyman}}]{brangwynne2009germline}%
  \BibitemOpen
  \bibfield  {author} {\bibinfo {author} {\bibfnamefont {C.~P.}\ \bibnamefont
  {Brangwynne}}, \bibinfo {author} {\bibfnamefont {C.~R.}\ \bibnamefont
  {Eckmann}}, \bibinfo {author} {\bibfnamefont {D.~S.}\ \bibnamefont
  {Courson}}, \bibinfo {author} {\bibfnamefont {A.}~\bibnamefont {Rybarska}},
  \bibinfo {author} {\bibfnamefont {C.}~\bibnamefont {Hoege}}, \bibinfo
  {author} {\bibfnamefont {J.}~\bibnamefont {Gharakhani}}, \bibinfo {author}
  {\bibfnamefont {F.}~\bibnamefont {J{\"u}licher}},\ and\ \bibinfo {author}
  {\bibfnamefont {A.~A.}\ \bibnamefont {Hyman}},\ }\bibfield  {title} {\bibinfo
  {title} {Germline p granules are liquid droplets that localize by controlled
  dissolution/condensation},\ }\href {https://doi.org/DOI:
  10.1126/science.1172046} {\bibfield  {journal} {\bibinfo  {journal}
  {Science}\ }\textbf {\bibinfo {volume} {324}},\ \bibinfo {pages} {1729}
  (\bibinfo {year} {2009})}\BibitemShut {NoStop}%
\bibitem [{\citenamefont {Hyman}\ \emph {et~al.}(2014)\citenamefont {Hyman},
  \citenamefont {Weber},\ and\ \citenamefont {J{\"u}licher}}]{hyman2014liquid}%
  \BibitemOpen
  \bibfield  {author} {\bibinfo {author} {\bibfnamefont {A.~A.}\ \bibnamefont
  {Hyman}}, \bibinfo {author} {\bibfnamefont {C.~A.}\ \bibnamefont {Weber}},\
  and\ \bibinfo {author} {\bibfnamefont {F.}~\bibnamefont {J{\"u}licher}},\
  }\bibfield  {title} {\bibinfo {title} {Liquid-liquid phase separation in
  biology},\ }\href
  {https://doi.org/https://doi.org/10.1146/annurev-cellbio-100913-013325}
  {\bibfield  {journal} {\bibinfo  {journal} {Annu. Rev. Cell Dev. Biol.}\
  }\textbf {\bibinfo {volume} {30}},\ \bibinfo {pages} {39} (\bibinfo {year}
  {2014})}\BibitemShut {NoStop}%
\bibitem [{\citenamefont {Weber}\ \emph {et~al.}(2019)\citenamefont {Weber},
  \citenamefont {Zwicker}, \citenamefont {J{\"u}licher},\ and\ \citenamefont
  {Lee}}]{weber2019physics}%
  \BibitemOpen
  \bibfield  {author} {\bibinfo {author} {\bibfnamefont {C.~A.}\ \bibnamefont
  {Weber}}, \bibinfo {author} {\bibfnamefont {D.}~\bibnamefont {Zwicker}},
  \bibinfo {author} {\bibfnamefont {F.}~\bibnamefont {J{\"u}licher}},\ and\
  \bibinfo {author} {\bibfnamefont {C.~F.}\ \bibnamefont {Lee}},\ }\bibfield
  {title} {\bibinfo {title} {Physics of active emulsions},\ }\href
  {https://doi.org/10.1088/1361-6633/ab052b} {\bibfield  {journal} {\bibinfo
  {journal} {Rep. Prog. Phys.}\ }\textbf {\bibinfo {volume} {82}},\ \bibinfo
  {pages} {064601} (\bibinfo {year} {2019})}\BibitemShut {NoStop}%
\bibitem [{\citenamefont {Halatek}\ and\ \citenamefont
  {Frey}(2012)}]{halatek2012highly}%
  \BibitemOpen
  \bibfield  {author} {\bibinfo {author} {\bibfnamefont {J.}~\bibnamefont
  {Halatek}}\ and\ \bibinfo {author} {\bibfnamefont {E.}~\bibnamefont {Frey}},\
  }\bibfield  {title} {\bibinfo {title} {Highly canalized mind transfer and
  mine sequestration explain the origin of robust mincde-protein dynamics},\
  }\href {https://doi.org/10.1016/j.celrep.2012.04.005} {\bibfield  {journal}
  {\bibinfo  {journal} {Cell Rep.}\ }\textbf {\bibinfo {volume} {1}},\ \bibinfo
  {pages} {741} (\bibinfo {year} {2012})}\BibitemShut {NoStop}%
\bibitem [{\citenamefont {Chardac}\ \emph {et~al.}(2025)\citenamefont
  {Chardac}, \citenamefont {Norton}, \citenamefont {Touboul},\ and\
  \citenamefont {Duclos}}]{chardac2025mind}%
  \BibitemOpen
  \bibfield  {author} {\bibinfo {author} {\bibfnamefont {A.}~\bibnamefont
  {Chardac}}, \bibinfo {author} {\bibfnamefont {M.~M.}\ \bibnamefont {Norton}},
  \bibinfo {author} {\bibfnamefont {J.}~\bibnamefont {Touboul}},\ and\ \bibinfo
  {author} {\bibfnamefont {G.}~\bibnamefont {Duclos}},\ }\bibfield  {title}
  {\bibinfo {title} {3d pattern formation of a protein--membrane suspension},\
  }\href {https://doi.org/https://doi.org/10.1073/pnas.2506401122} {\bibfield
  {journal} {\bibinfo  {journal} {PNAS}\ }\textbf {\bibinfo {volume} {122}},\
  \bibinfo {pages} {e2506401122} (\bibinfo {year} {2025})}\BibitemShut
  {NoStop}%
\bibitem [{\citenamefont {Brauns}\ \emph {et~al.}(2021)\citenamefont {Brauns},
  \citenamefont {Weyer}, \citenamefont {Halatek}, \citenamefont {Yoon},\ and\
  \citenamefont {Frey}}]{brauns2021wavelength}%
  \BibitemOpen
  \bibfield  {author} {\bibinfo {author} {\bibfnamefont {F.}~\bibnamefont
  {Brauns}}, \bibinfo {author} {\bibfnamefont {H.}~\bibnamefont {Weyer}},
  \bibinfo {author} {\bibfnamefont {J.}~\bibnamefont {Halatek}}, \bibinfo
  {author} {\bibfnamefont {J.}~\bibnamefont {Yoon}},\ and\ \bibinfo {author}
  {\bibfnamefont {E.}~\bibnamefont {Frey}},\ }\bibfield  {title} {\bibinfo
  {title} {Wavelength selection by interrupted coarsening in reaction-diffusion
  systems},\ }\href
  {https://doi.org/https://doi.org/10.1103/PhysRevLett.126.104101} {\bibfield
  {journal} {\bibinfo  {journal} {Phys. Rev. Lett.}\ }\textbf {\bibinfo
  {volume} {126}},\ \bibinfo {pages} {104101} (\bibinfo {year}
  {2021})}\BibitemShut {NoStop}%
\bibitem [{\citenamefont {Cates}\ and\ \citenamefont
  {Nardini}(2025)}]{cates2025active}%
  \BibitemOpen
  \bibfield  {author} {\bibinfo {author} {\bibfnamefont {M.~E.}\ \bibnamefont
  {Cates}}\ and\ \bibinfo {author} {\bibfnamefont {C.}~\bibnamefont
  {Nardini}},\ }\bibfield  {title} {\bibinfo {title} {Active phase separation:
  new phenomenology from non-equilibrium physics},\ }\href
  {https://doi.org/0.1088/1361-6633/add278} {\bibfield  {journal} {\bibinfo
  {journal} {Rep. Prog. Phys.}\ }\textbf {\bibinfo {volume} {88}},\ \bibinfo
  {pages} {056601} (\bibinfo {year} {2025})}\BibitemShut {NoStop}%
\bibitem [{\citenamefont {Tjhung}\ \emph {et~al.}(2018)\citenamefont {Tjhung},
  \citenamefont {Nardini},\ and\ \citenamefont {Cates}}]{tjhung2018cluster}%
  \BibitemOpen
  \bibfield  {author} {\bibinfo {author} {\bibfnamefont {E.}~\bibnamefont
  {Tjhung}}, \bibinfo {author} {\bibfnamefont {C.}~\bibnamefont {Nardini}},\
  and\ \bibinfo {author} {\bibfnamefont {M.~E.}\ \bibnamefont {Cates}},\
  }\bibfield  {title} {\bibinfo {title} {Cluster phases and bubbly phase
  separation in active fluids: Reversal of the ostwald process},\ }\href
  {https://doi.org/https://doi.org/10.1103/PhysRevX.8.031080} {\bibfield
  {journal} {\bibinfo  {journal} {Phys. Rev. X}\ }\textbf {\bibinfo {volume}
  {8}},\ \bibinfo {pages} {031080} (\bibinfo {year} {2018})}\BibitemShut
  {NoStop}%
\bibitem [{\citenamefont {Fausti}\ \emph {et~al.}(2021)\citenamefont {Fausti},
  \citenamefont {Tjhung}, \citenamefont {Cates},\ and\ \citenamefont
  {Nardini}}]{fausti2021capillary}%
  \BibitemOpen
  \bibfield  {author} {\bibinfo {author} {\bibfnamefont {G.}~\bibnamefont
  {Fausti}}, \bibinfo {author} {\bibfnamefont {E.}~\bibnamefont {Tjhung}},
  \bibinfo {author} {\bibfnamefont {M.~E.}\ \bibnamefont {Cates}},\ and\
  \bibinfo {author} {\bibfnamefont {C.}~\bibnamefont {Nardini}},\ }\bibfield
  {title} {\bibinfo {title} {Capillary interfacial tension in active phase
  separation},\ }\href
  {https://doi.org/https://doi.org/10.1103/PhysRevLett.127.068001} {\bibfield
  {journal} {\bibinfo  {journal} {Phys. Rev. Lett.}\ }\textbf {\bibinfo
  {volume} {127}},\ \bibinfo {pages} {068001} (\bibinfo {year}
  {2021})}\BibitemShut {NoStop}%
\bibitem [{\citenamefont {Bergmann}\ \emph {et~al.}(2018)\citenamefont
  {Bergmann}, \citenamefont {Rapp},\ and\ \citenamefont
  {Zimmermann}}]{bergmann2018active}%
  \BibitemOpen
  \bibfield  {author} {\bibinfo {author} {\bibfnamefont {F.}~\bibnamefont
  {Bergmann}}, \bibinfo {author} {\bibfnamefont {L.}~\bibnamefont {Rapp}},\
  and\ \bibinfo {author} {\bibfnamefont {W.}~\bibnamefont {Zimmermann}},\
  }\bibfield  {title} {\bibinfo {title} {Active phase separation: A universal
  approach},\ }\href
  {https://doi.org/https://doi.org/10.1103/PhysRevE.98.020603} {\bibfield
  {journal} {\bibinfo  {journal} {Phys. Rev. E}\ }\textbf {\bibinfo {volume}
  {98}},\ \bibinfo {pages} {020603} (\bibinfo {year} {2018})}\BibitemShut
  {NoStop}%
\bibitem [{\citenamefont {Bergmann}\ and\ \citenamefont
  {Zimmermann}(2019)}]{bergmann2019system}%
  \BibitemOpen
  \bibfield  {author} {\bibinfo {author} {\bibfnamefont {F.}~\bibnamefont
  {Bergmann}}\ and\ \bibinfo {author} {\bibfnamefont {W.}~\bibnamefont
  {Zimmermann}},\ }\bibfield  {title} {\bibinfo {title} {On system-spanning
  demixing properties of cell polarization},\ }\href
  {https://doi.org/https://doi.org/10.1371/journal.pone.0218328} {\bibfield
  {journal} {\bibinfo  {journal} {PloS one}\ }\textbf {\bibinfo {volume}
  {14}},\ \bibinfo {pages} {e0218328} (\bibinfo {year} {2019})}\BibitemShut
  {NoStop}%
\bibitem [{\citenamefont {Robinson}\ \emph {et~al.}(2025)\citenamefont
  {Robinson}, \citenamefont {Machon},\ and\ \citenamefont
  {Speck}}]{robinson2025universal}%
  \BibitemOpen
  \bibfield  {author} {\bibinfo {author} {\bibfnamefont {J.~F.}\ \bibnamefont
  {Robinson}}, \bibinfo {author} {\bibfnamefont {T.}~\bibnamefont {Machon}},\
  and\ \bibinfo {author} {\bibfnamefont {T.}~\bibnamefont {Speck}},\ }\bibfield
   {title} {\bibinfo {title} {Universal limiting behavior of reaction-diffusion
  systems with conservation laws},\ }\href
  {https://doi.org/https://doi.org/10.1103/1bdc-1bjb} {\bibfield  {journal}
  {\bibinfo  {journal} {Phys. Rev. E}\ }\textbf {\bibinfo {volume} {111}},\
  \bibinfo {pages} {065417} (\bibinfo {year} {2025})}\BibitemShut {NoStop}%
\bibitem [{\citenamefont {Toffenetti}\ \emph {et~al.}(2026)\citenamefont
  {Toffenetti}, \citenamefont {Nettuno}, \citenamefont {Weyer},\ and\
  \citenamefont {Frey}}]{toffenetti2026active}%
  \BibitemOpen
  \bibfield  {author} {\bibinfo {author} {\bibfnamefont {D.}~\bibnamefont
  {Toffenetti}}, \bibinfo {author} {\bibfnamefont {B.}~\bibnamefont {Nettuno}},
  \bibinfo {author} {\bibfnamefont {H.}~\bibnamefont {Weyer}},\ and\ \bibinfo
  {author} {\bibfnamefont {E.}~\bibnamefont {Frey}},\ }\bibfield  {title}
  {\bibinfo {title} {Active model b- from mass-conserving reaction-diffusion
  systems},\ }\bibfield  {journal} {\bibinfo  {journal} {arXiv preprint
  arXiv:2605.15903}\ }\href
  {https://doi.org/https://doi.org/10.48550/arXiv.2605.15903}
  {https://doi.org/10.48550/arXiv.2605.15903} (\bibinfo {year}
  {2026})\BibitemShut {NoStop}%
\bibitem [{\citenamefont {Ilker}\ \emph {et~al.}(2026)\citenamefont {Ilker},
  \citenamefont {Laxhuber}, \citenamefont {Joanny},\ and\ \citenamefont
  {J{\"u}licher}}]{ilker2026hydrodynamic}%
  \BibitemOpen
  \bibfield  {author} {\bibinfo {author} {\bibfnamefont {E.}~\bibnamefont
  {Ilker}}, \bibinfo {author} {\bibfnamefont {K.}~\bibnamefont {Laxhuber}},
  \bibinfo {author} {\bibfnamefont {J.-F.}\ \bibnamefont {Joanny}},\ and\
  \bibinfo {author} {\bibfnamefont {F.}~\bibnamefont {J{\"u}licher}},\
  }\bibfield  {title} {\bibinfo {title} {Hydrodynamic theory of chemically
  active emulsions},\ }\href {https://arxiv.org/pdf/2604.17539} {\bibfield
  {journal} {\bibinfo  {journal} {arXiv preprint arXiv:2604.17539}\ } (\bibinfo
  {year} {2026})}\BibitemShut {NoStop}%
\bibitem [{\citenamefont {Golestanian}(2022)}]{Golestanian2022a}%
  \BibitemOpen
  \bibfield  {author} {\bibinfo {author} {\bibfnamefont {R.}~\bibnamefont
  {Golestanian}},\ }\bibfield  {title} {\bibinfo {title} {Phoretic active
  matter},\ }in\ \href {https://doi.org/10.1093/oso/9780192858313.003.0008}
  {\emph {\bibinfo {booktitle} {{Active Matter and Nonequilibrium Statistical
  Physics. Lecture Notes of the Les Houches Summer School}}}}\ (\bibinfo
  {publisher} {Oxford University Press},\ \bibinfo {address} {UK},\ \bibinfo
  {year} {2022})\ pp.\ \bibinfo {pages} {230--293}\BibitemShut {NoStop}%
\bibitem [{\citenamefont {Roussel}\ and\ \citenamefont
  {Roussel}(2004)}]{roussel2004reaction}%
  \BibitemOpen
  \bibfield  {author} {\bibinfo {author} {\bibfnamefont {C.~J.}\ \bibnamefont
  {Roussel}}\ and\ \bibinfo {author} {\bibfnamefont {M.~R.}\ \bibnamefont
  {Roussel}},\ }\bibfield  {title} {\bibinfo {title} {Reaction--diffusion
  models of development with state-dependent chemical diffusion coefficients},\
  }\href {https://doi.org/https://doi.org/10.1016/j.pbiomolbio.2004.03.001}
  {\bibfield  {journal} {\bibinfo  {journal} {Prog. Biophys. Mol. Biol.}\
  }\textbf {\bibinfo {volume} {86}},\ \bibinfo {pages} {113} (\bibinfo {year}
  {2004})}\BibitemShut {NoStop}%
\bibitem [{\citenamefont {Polanowski}\ and\ \citenamefont
  {Koza}(2006)}]{polanowski2006reaction}%
  \BibitemOpen
  \bibfield  {author} {\bibinfo {author} {\bibfnamefont {P.}~\bibnamefont
  {Polanowski}}\ and\ \bibinfo {author} {\bibfnamefont {Z.}~\bibnamefont
  {Koza}},\ }\bibfield  {title} {\bibinfo {title} {Reaction-diffusion fronts in
  systems with concentration-dependent diffusivities},\ }\href
  {https://doi.org/10.1103/PhysRevE.74.036103} {\bibfield  {journal} {\bibinfo
  {journal} {Phys. Rev. E}\ }\textbf {\bibinfo {volume} {74}},\ \bibinfo
  {pages} {036103} (\bibinfo {year} {2006})}\BibitemShut {NoStop}%
\bibitem [{\citenamefont {Roussel}\ and\ \citenamefont
  {Wang}(2004)}]{roussel2004pattern}%
  \BibitemOpen
  \bibfield  {author} {\bibinfo {author} {\bibfnamefont {M.~R.}\ \bibnamefont
  {Roussel}}\ and\ \bibinfo {author} {\bibfnamefont {J.}~\bibnamefont {Wang}},\
  }\bibfield  {title} {\bibinfo {title} {Pattern formation in excitable media
  with concentration-dependent diffusivities},\ }\href
  {https://doi.org/https://doi.org/10.1063/1.1689640} {\bibfield  {journal}
  {\bibinfo  {journal} {J. Chem. Phys.}\ }\textbf {\bibinfo {volume} {120}},\
  \bibinfo {pages} {8079} (\bibinfo {year} {2004})}\BibitemShut {NoStop}%
\bibitem [{\citenamefont {De~Groot}\ and\ \citenamefont
  {Mazur}(2013)}]{de2013non}%
  \BibitemOpen
  \bibfield  {author} {\bibinfo {author} {\bibfnamefont {S.~R.}\ \bibnamefont
  {De~Groot}}\ and\ \bibinfo {author} {\bibfnamefont {P.}~\bibnamefont
  {Mazur}},\ }\href@noop {} {\emph {\bibinfo {title} {Non-equilibrium
  thermodynamics}}}\ (\bibinfo  {publisher} {Courier Corporation},\ \bibinfo
  {year} {2013})\BibitemShut {NoStop}%
\bibitem [{\citenamefont {Avanzini}\ \emph {et~al.}(2024)\citenamefont
  {Avanzini}, \citenamefont {Aslyamov}, \citenamefont {Fodor},\ and\
  \citenamefont {Esposito}}]{avanzini2024nonequilibrium}%
  \BibitemOpen
  \bibfield  {author} {\bibinfo {author} {\bibfnamefont {F.}~\bibnamefont
  {Avanzini}}, \bibinfo {author} {\bibfnamefont {T.}~\bibnamefont {Aslyamov}},
  \bibinfo {author} {\bibfnamefont {{\'E}.}~\bibnamefont {Fodor}},\ and\
  \bibinfo {author} {\bibfnamefont {M.}~\bibnamefont {Esposito}},\ }\bibfield
  {title} {\bibinfo {title} {Nonequilibrium thermodynamics of non-ideal
  reaction--diffusion systems: Implications for active self-organization},\
  }\href {https://doi.org/10.1063/5.0231520} {\bibfield  {journal} {\bibinfo
  {journal} {J. Chem. Phys.}\ }\textbf {\bibinfo {volume} {161}} (\bibinfo
  {year} {2024})}\BibitemShut {NoStop}%
\bibitem [{\citenamefont {Akaberian}\ \emph {et~al.}(2023)\citenamefont
  {Akaberian}, \citenamefont {Thewes}, \citenamefont {Sollich},\ and\
  \citenamefont {Kr{\"u}ger}}]{akaberian2023nonequilibrium}%
  \BibitemOpen
  \bibfield  {author} {\bibinfo {author} {\bibfnamefont {M.}~\bibnamefont
  {Akaberian}}, \bibinfo {author} {\bibfnamefont {F.~C.}\ \bibnamefont
  {Thewes}}, \bibinfo {author} {\bibfnamefont {P.}~\bibnamefont {Sollich}},\
  and\ \bibinfo {author} {\bibfnamefont {M.}~\bibnamefont {Kr{\"u}ger}},\
  }\bibfield  {title} {\bibinfo {title} {Nonequilibrium mixture dynamics: A
  model for mobilities and its consequences},\ }\href
  {https://doi.org/10.1063/5.0147206} {\bibfield  {journal} {\bibinfo
  {journal} {J. Chem. Phys.}\ }\textbf {\bibinfo {volume} {158}} (\bibinfo
  {year} {2023})}\BibitemShut {NoStop}%
\bibitem [{SM()}]{SM}%
  \BibitemOpen
  \bibinfo {note} {See Supplemental Material: It includes a detailed discussion
  on the derivation of the mobility matrix starting from discrete jump
  processes, on the derivation of reactive and conservative noise of Eq.(1) and
  numerical details of Fig. 2. It also includes Ref
  \cite{gardiner2009stochastic, johnsrud2025efficient}}\BibitemShut {NoStop}%
\bibitem [{\citenamefont {Gardiner}(2009)}]{gardiner2009stochastic}%
  \BibitemOpen
  \bibfield  {author} {\bibinfo {author} {\bibfnamefont {C.}~\bibnamefont
  {Gardiner}},\ }\href@noop {} {\emph {\bibinfo {title} {Stochastic
  methods}}},\ Vol.~\bibinfo {volume} {4}\ (\bibinfo  {publisher} {Springer
  Berlin Heidelberg},\ \bibinfo {year} {2009})\BibitemShut {NoStop}%
\bibitem [{\citenamefont {Dean}(1996)}]{dean1996langevin}%
  \BibitemOpen
  \bibfield  {author} {\bibinfo {author} {\bibfnamefont {D.~S.}\ \bibnamefont
  {Dean}},\ }\bibfield  {title} {\bibinfo {title} {Langevin equation for the
  density of a system of interacting langevin processes},\ }\href
  {https://iopscience.iop.org/article/10.1088/0305-4470/29/24/001} {\bibfield
  {journal} {\bibinfo  {journal} {J. Phys. A: Math. Gen}\ }\textbf {\bibinfo
  {volume} {29}},\ \bibinfo {pages} {L613} (\bibinfo {year}
  {1996})}\BibitemShut {NoStop}%
\bibitem [{\citenamefont {Mori}\ \emph {et~al.}(2008)\citenamefont {Mori},
  \citenamefont {Jilkine},\ and\ \citenamefont
  {Edelstein-Keshet}}]{mori2008wave}%
  \BibitemOpen
  \bibfield  {author} {\bibinfo {author} {\bibfnamefont {Y.}~\bibnamefont
  {Mori}}, \bibinfo {author} {\bibfnamefont {A.}~\bibnamefont {Jilkine}},\ and\
  \bibinfo {author} {\bibfnamefont {L.}~\bibnamefont {Edelstein-Keshet}},\
  }\bibfield  {title} {\bibinfo {title} {Wave-pinning and cell polarity from a
  bistable reaction-diffusion system},\ }\href {10.1529/biophysj.107.120824}
  {\bibfield  {journal} {\bibinfo  {journal} {Biophys. J.}\ }\textbf {\bibinfo
  {volume} {94}},\ \bibinfo {pages} {3684} (\bibinfo {year}
  {2008})}\BibitemShut {NoStop}%
\bibitem [{\citenamefont {Burekovi{\'c}}\ \emph {et~al.}(2026)\citenamefont
  {Burekovi{\'c}}, \citenamefont {De~Luca}, \citenamefont {Cates},\ and\
  \citenamefont {Nardini}}]{burekovic2026active}%
  \BibitemOpen
  \bibfield  {author} {\bibinfo {author} {\bibfnamefont {S.}~\bibnamefont
  {Burekovi{\'c}}}, \bibinfo {author} {\bibfnamefont {F.}~\bibnamefont
  {De~Luca}}, \bibinfo {author} {\bibfnamefont {M.~E.}\ \bibnamefont {Cates}},\
  and\ \bibinfo {author} {\bibfnamefont {C.}~\bibnamefont {Nardini}},\
  }\bibfield  {title} {\bibinfo {title} {Active cahn-hilliard theory for
  nonequilibrium phase separation: Quantitative macroscopic predictions and a
  microscopic derivation},\ }\href {https://doi.org/10.1103/n155-s73b}
  {\bibfield  {journal} {\bibinfo  {journal} {Phys. Rev. Res.}\ }\textbf
  {\bibinfo {volume} {8}},\ \bibinfo {pages} {033141} (\bibinfo {year}
  {2026})}\BibitemShut {NoStop}%
\bibitem [{\citenamefont {Johnsrud}\ and\ \citenamefont
  {Rana}(2025)}]{johnsrud2025efficient}%
  \BibitemOpen
  \bibfield  {author} {\bibinfo {author} {\bibfnamefont {M.~K.}\ \bibnamefont
  {Johnsrud}}\ and\ \bibinfo {author} {\bibfnamefont {N.}~\bibnamefont
  {Rana}},\ }\bibfield  {title} {\bibinfo {title} {Efficient pseudo-spectral
  algorithms for statistical field theories},\ }\href
  {https://arxiv.org/pdf/2511.22510} {\bibfield  {journal} {\bibinfo  {journal}
  {arXiv preprint arXiv:2511.22510}\ } (\bibinfo {year} {2025})}\BibitemShut
  {NoStop}%
\end{thebibliography}%

\section{End Matter}

\textit{Coefficients of Eq.\eqref{eq:nonlinearphi}.}
\begin{align}
 b_0 &= \frac{1}{(R_p-R_s)^2}\Bigl[D_s^{ss} R_p^2 + D_p^{pp} R_s^2 - (D_s^{pp}+D_p^{ss}) R_s R_p\Bigr], \nonumber \\ 
 b_1 &= \frac{1}{(R_p-R_s)} \Bigl[
R_s(D_s^{pp}-D_p^{pp}) + R_p (D_p^{ss} - D_s^{ss})\Bigr], \\
 b_2 &=\frac{1}{(R_p-R_s)}
\Bigl[
R_s(D_p^{ss}-D_p^{pp}) + R_p (D_s^{pp} - D_s^{ss})
\Bigr], \nonumber \\
 b_3 &= \frac{1}{2 (R_p-R_s)^3}\Bigl[ R_p^3 D_{ss}^{ss} - R_s^3 D_{pp}^{pp} -R_s R_p^2 ( D_{ss}^{pp} + 2 D_{ss}^{ps} + \nonumber \\ 
& \qquad + R_p R_s^2 ( D_{pp}^{ss} + 2 D_{pp}^{ps}) \Bigr] \nonumber .
\end{align} 
\newline 

\textit{Coefficients of Eq.\eqref{eq:nonlineta}.}
\begin{align}
\nonumber
 b_4 &= - \frac{R_p R_s}{2 (R_p-R_s)^2} b_2, \\ 
 a_2 &= \frac{1}{2 (R_p-R_s)^2}(R_s^2 R_{pp} +R_p^2 R_{ss} - 2 R_s R_p R_{sp}), \\ \nonumber
 a_3 &= \frac{1}{6 (R_p-R_s)^3} \Bigl[ R_s^2 ( 3 R_p R_{spp} - R_s R_{ppp}) + R_p^2 (R_{sss} -3 R_s R_{ssp}) \Bigr], \\ \nonumber
 a_4 &= \frac{1}{(R_p-R_s)} \Bigl[ (R_p+R_s) R_{sp}- R_p R_{ss} -R_s R_{pp}\Bigr]  .
\end{align}
\end{document}